# Wettability Induced Crack Dynamics and Morphology

*Udita Uday Ghosh[1], Monojit Chakraborty[1], Aditya Bikram Bhandari[1], Suman Chakraborty[2] and*

*Sunando DasGupta[1]\**

[1]*Department of Chemical Engineering, Indian Institute of Technology, Kharagpur 721302*

[2]*Department of Mechanical Engineering, Indian Institute of Technology, Kharagpur 721302*

\*Corresponding author

Email      sunando@che.iitkgp.ernet.in; sunando.dasgupta@gmail.com

Ph:      +91 - 3222 - 283922

## Graphical Abstract

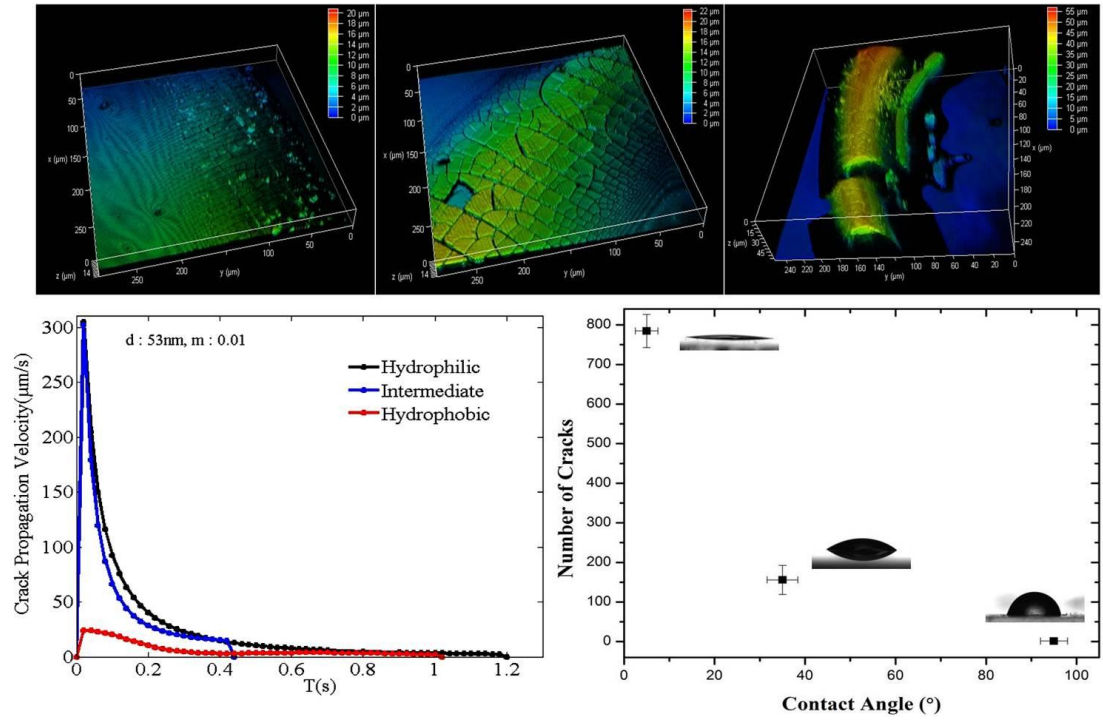

**Abstract**

Substrate wettability alteration induced control over crack formation process in thin colloidal films has been addressed in the present study. Colloidal nanosuspension (53nm, mean particle diameter) droplets have been subjected to natural drying to outline the effects of substrate surface energies over the dry-out characteristics with emphasis on crack dynamics, crack morphology and underlying particle arrangements. Experimental findings indicate that number of cracks formed decreases with with increase in substrate hydrophobicity. These physical phenomena have been explained based on the magnitude of stress dissipation incurred by the substrate. DLVO predictions are also found to be in tune with the reported experimental investigations.

# 1. Introduction

Microdroplets of colloidal suspension on evaporation produce characteristic patterns comprising of self assembled layers of colloidal particles. Various methods that have been devised to produce materials using directed self-assembly of colloidal particles include sedimentation[1], evaporation[2], adsorption or layer-by-layer deposition[3], external force field[4], surface[5] and droplet[6] templating. These assemblies are of great importance in the upcoming domains of photonics, electronics, sensing, and drug delivery[7]. It has also been reported that self-assembled particles may function as microchannels[8]. Radially aligned and Periodic array of microchannels have been obtained by altering the mode of evaporation from 'free' to 'constrained'. Undoubtedly, Microchannels form and serve as a basic unit in many microfluidic devices [9]and thereby self-assembly opens up a new avenue in fabrication of microchannels. On similar lines, biological fluid droplets present distinctly different patterns under diseased and normal conditions. This may serve as an indicative test for disease detection[10]. A new term 'Nanochromatography' has also been coined as a complement to the otherwise well known chromatography technique, commonly used to separate solutes from a mixture. Nanochromatography on the other hand exploits the size dependency of 'coffee-ring effect' at the rim, precisely at the contact line of the evaporating droplet for separation of proteins, micro organisms, and mammalian cells alike[11]. Characteristic patterns formed after evaporation of the solvent may also serve as an evolving technique in patterning technology. These patterns are mini-scale representation of the craquelures observed in paintings which reveal pigment composition and determine the longevity of paints[12]. Thereby, the entire process of drying of colloidal suspensions has been of considerable importance in the industrial and academic scenario alike owing to its widespread applications across domains ranging from fabrication of durable, economic and environment friendly coatings[13], photonic crystals/opals[14] to materials formed by directed assembly of micro and nano-particles[15]. It is still a challenging task to obtain ordered structures of desired functionalities in spite of the advent of techniques of drying mediated assembly of colloidal particles[16].

Solvent evaporation is the primary step in these colloidal assembly formation. It is governed by the mode of drying and these modes can be classified under three categories-lateral, isotropic[17](uniform drying) and directional. Lateral drying involves coating the substrate with the colloidal dispersion[18] and directional drying constraints the film to be de-moist from a single direction only[19]. Subsequent steps involve compaction of solvent devoid particles leading to an increase in inherent film capillary pressure. The concluding step is the collapse of the film manifested in the form of film fracture (cracks). It is well known that crack formation is highly detrimental to the potential usage of the final product. This not only requires control over the fabrication process but also necessitates deceleration of crack propagation. An ideal situation would be the 'surface modified substrates resistant to fracture'. It becomes inevitable at this juncture to delve into the general hypotheses put forth to explain the underlying mechanisms of crack formation process.

Dufresne et.al[22] suggested that crack growth is directly dependant on the movement of the compaction front that is governed by the rate of fluid evaporation and fluid refilling rate at the air-fluid interface due to the capillary tension created by evaporation. Thereby the length scale of the compaction front demarcates the two observable regimes namely evaporation-limited and flow-limited. Routh et.al [23] proposed a scaling of the crack spacing with film thickness as a function of the generated capillary pressure and held the capillary stresses responsible for the failure of the dried colloidal film. Particle size dependency was also probed and reported in affirmation[24,25].

The resultant crack morphology based on the mechanisms enlisted above is an important aspect of crack characterization. It is a function of the drying conditions, particle properties (size, shape, charge) and substrate characteristics. Stripe patterns were obtained on evaporation of nano-suspension droplets (PS,144nm) over glass surface[26]. Theoretical analysis suggested that the stripe patterns were formed as a result of competition between the droplet surface tension and frictional force at the contact line. On the other hand concentric (multiple) rings were observed, on drying latex films (comprising of micrometre sized particles), over glass surfaces. This difference in crack morphology (stripe and ring) has been attributed to the ease of nucleation in the microsuspensions as compared to nano-suspensions. Nanosuspensions present a characteristic mesh of arches intertwined at times with each other at the arch-edges in contrast to the parallel crack observed in the case of microsuspensions[27]. Concentric, circular cracks were obtained on drying silica nano-suspensions, confirming the Xia-Hutchinson model hypothesis that a pre-existing flaw loop in the film is responsible for crack initiation as well as it's propagation[28]. Mechanism of crack formation provides the background necessary to tackle crack arrest. Several attempts have been directed towards resolving this problem which includes usage of soft particles[13] or coating hard particles with soft shells[29], addition of polymeric plasticizers[30] and employment of sequential deposition of multiple layers[31]. Experimental reports of addition of emulsion provide evidence that can crack formation can not only be decelerated but it may be even possible to eliminate it above a critical particle concentration.

However, the methods proposed so far have primarily focus on altering, the suspension composition, mechanical properties of the gel[32] or drying kinetics[33]. We may now direct our attention towards the role that the substrate plays during the drying phenomenon and consequentially the crack formation process[34]. This has been probed further by replacing the solid substrate with a liquid mercury surface for drying 400 nm alumina particles[24]. Recently, experimental outcomes of crack formation studies on compliant elastomer substrates with varying substrate elastic modulus indicated that the characteristic length scales of cracks were inversely proportional to substrate elasticity[25]. This further highlights the importance of in-plane constraints/tensile stresses in the process of crack formation[35]. It has also been reported that the crack shape undergoes change with variation in surface functional group attached to the substrate and this has been attributed to the substrate-particle interactions[36].

However, the role of surface characteristics of the substrate in the crack formation process has not yet been explored to our knowledge. Surface characteristics primarily include properties like surface roughness, wettability, elasticity, permeability etc. Thereby, this study expounds the effect of substrate surface properties, in particular 'substrate wettability' on the crack dynamics and the resultant crack morphology/pattern .The in-plane constraints/tensile stresses may be manipulated by altering the substrate wettability characteristics. Such a manipulation requires a quantitative and qualitative study of the effect of substrate surface characteristics. Micro-droplets of nanosuspensions have been dried to record the patterns and outline the crack dynamics over substrates with characteristic wettability.

## 2. Materials and Methods

### 2.1 Materials preparation

*A. Preparation of substrate and its characterisation*

Glass slides (BLUESTAR, Polar Industrial Corporation, INDIA) were used as substrates. The cleaning protocol comprised of ultrasonication in acetone followed by ultrasonication in deionized water for a duration of 10minutes each to remove dust and organic contaminants. The substrates were classified into three major categories based on their wettability:

1. *Hydrophilic:* Cleaned glass slides plasma treated for 60seconds.

2. *Intermediate:* Cleaned glass slides without any pre- treatment.

3. *Hydrophobic:* Substrates were coated with a layer of PDMS to render them hydrophobic. The base and the cross-linker (SYLGARD 184, obtained from Dow Corning) were mixed in the weight ratio of 10:1. The mixture was then degassed to remove any air-bubbles formed during mixing. The glass slide was coated with the mixture using a spin coater. The coating process comprises of a slow spreading step at 500 rpm for 30 seconds followed by 5000 rpm for 70 seconds. The formed PDMS layer was cured by placing it overnight in a hot air oven at 95 °C. The thickness of the layer was measured using a profilometer and was found to be ~13 μm.

Sessile droplets of DI water and colloidal dispersion (1µl volume and 1(w/w) %) were placed on three types substrates differing in wettability. Contact angles were evaluated using a Goniometer (Ramehart, Germany) and corresponding contact angles have been enlisted in Table I.

TABLE I  Characterisation of Substrate Wettability

| Index | Substrate | Mean Equilibrium Contact Angle ($\theta \pm 2$) ($^0$) | |
|---|---|---|---|
| | | DI water (a) | Colloidal droplet (b) |
| S1 | Hydrophilic | 5 | 5 |
| S2 | Intermediate | 35 | 36 |
| S3 | Hydrophobic | 95 | 96 |

*B. Preparation of Colloidal suspension*

Colloidal system comprised of aqueous suspensions of polystyrene latex beads with particle diameter of 0.053µm obtained from Sigma-Aldrich (Solid content 10(w/w)%). Deionized water (Milli-Q, 99% pure) was used to dilute the colloidal suspension to the required concentration (1(w/w) %).These nanosuspensions were sonicated for 10 minutes prior to each experiment to ensure homogeneity.

**2.2 Experimental Procedure**

Colloidal droplets (1µl volume and 1 (w/w)%) were placed on the characterized substrates and were allowed to dry by natural evaporation at ambient conditions The temperature and humidity conditions were maintained at ($25^0$C) and 40% relative humidity. The solvent gradually evaporated sweeping the particles to the droplet periphery. The drying front was found to undergo a transformation in to a thick region formed by particle compaction, easily distinguishable from the rest of the droplet by the optical contrast. This was followed by crack initiation in these thin colloidal films.

**2.2.1. Optical microscopy**

A microscope (LEICA DM5200) was operated in the incident bright field mode and focussed at the periphery of the sessile colloidal droplet. Time-lapse videos were recorded featuring the stages of crack formation with 50X objective for hydrophilic(S1) and intermediate(S2) substrates and 10X objective for hydrophobic(S3) substrates. Images were extracted from these videos and analyzed to evaluate the advancement of crack tip position with time. These spatial curves were fitted to standard curves to determine instantaneous velocity of the crack tip. Experiments over each substrate were repeated and the results were found to be reproducible.

The 50X objective imposed a constraint on the interrogation area limiting the visibility on S1and S2substrates during live recording of the crack formation process. However, the entire perimeter of the dried films gets filled cracks. Thereby for determining the total number of cracks for each dried film a manual procedure has been adopted. An enlarged image of each dried colloidal film has been taken by merging the individual interrogation areas, exposing the entire film to enumerate the cracks.

**2.2.2. Confocal Microscopy**

Dried colloidal films were viewed under a Confocal microscope (LEICA TCS SP5) to examine the morphology of each type of crack pattern. The *crack opening δ* and *the distance along the crack from its tip r* were evaluated using magnified images of individual cracks. Thicknesses of the dried films were evaluated in the RT(reflection-transmission) mode by forming z-stacks of each film [37].

**2.2.3. Atomic Force Microscopy (AFM) and Scanning Electron Microscopy (SEM) Imaging**

A digital AGILENT (Nanonics Model no.5100) atomic force microscope equipped with a PicoView software has been used to image the region in the proximity of the crack. The AFM used herein has a silicon cantilever and has been operated in the intermittent contact imaging mode.

Scanning electron microscopy (NovaNANO FESEM) has also been used to probe the particle arrangements close to fracture zone. It has been operated in the high pressure vaccum mode.

## 3. Theory

Derjaguin-Landau-Verwey-Overbeek(DLVO) theory has been invoked to account for the colloidal interactions in the evaporating colloidal droplets[38]. Colloidal particles in the present analysis are assumed to be hard spheres relieving the subsequent analysis of the consideration of possible deformation caused by particle compression during drying. These non-deformable hard spheres are assumed to interact with each other on a one to one basis with negligible interaction with its neighbours. It has been postulated that such non-pairwise interaction may be neglected if $kr \geq 5$, where $k$ is the reciprocal of the Debye length and $r$ is the particle radius[39]. This condition stands valid for the current study, thereby the expressions for forces account for pair-wise interactions only.

The total DLVO force for the colloidal system comprises of the Electrostatic and Van der Waals forces.

$$F_{DLVO} = F_{VdW} + F_{el} \quad \ldots \ldots (1)$$
$$= (F_{substrate-particle} + F_{particle-particle})_{VdW} + (F_{substrate-particle} + F_{particle-particle})_{el}$$
$$= (F_{substrate-particle})_{VdW} + (F_{particle-particle})_{VdW} + (F_{substrate-particle})_{el} + (F_{particle-particle})_{el} \quad \ldots \ldots (2)$$

Details of the individual force components are given below,

**A. *Van der Waals forces***

$$F_{VdW, particle-substrate} = \frac{2nA_{123}r^3}{3z^2(z+2r)^2} \quad \ldots (3)$$

$$F_{VdW, particle-particle} = \frac{nA_{131}r}{12z^2} \quad \ldots (4)$$

Van der Waals forces acting between particle -substrate and particle -particle are expressed in equations (3) and (4) where, $r$ is the particle radius, $n$ has been assumed to be the maximum number of particles in closed packed crystalline arrangement formed as a result of crack front opening and subsequent compression of particles and $z$ is the minimum separation distance between the particles and substrate. Although substrate wettability alteration may affect $n$ and $z$ but it is not feasible to discern this alteration with the aid of current experimental investigations.

The Hamaker constant designated as $A_{123}$[40] is for a system comprising of colloidal particles $(A_{11})$ in an intervening fluid medium $(A_{33})$ between the particles and substrate $(A_{22})$. Similarly, $A_{131}$ is the Hamaker constant for the particle-particle interaction in fluid medium ((5a) and (5b)).

$$A_{123} = \left(\sqrt{A_{11}} - \sqrt{A_{33}}\right)\left(\sqrt{A_{22}} - \sqrt{A_{33}}\right) \quad \ldots(5a)$$

$$A_{131} = (\sqrt{A_{11}} - \sqrt{A_{33}})^2 \quad \ldots\ldots\ldots\ldots\ldots\ldots(5b)$$

### B. *Electrostatic forces*

Inter-particle and particle-substrate electrostatic forces are shown in equation (6) and (7),

$$F_{el,particle-substrate} = -2n\pi\varepsilon rk \frac{[\phi_1^2 + \phi_2^2 - 2\phi_1\phi_2 \exp(kz)]}{[\exp(2kz) - 1]} \quad \ldots(6)$$

$$F_{el,particle-particle} = -\frac{\varepsilon r\phi_1^2}{4} \frac{2k\exp(kz)}{\exp(2kz) + 1} \quad \ldots(7)$$

where, $\varepsilon$ is the permittivity of the fluid medium (water), $k$ refers to the reciprocal of the Debye length, $\phi_1$ and $\phi_2$ are the surface potentials of the particle and the substrate respectively. Surface potentials equivalent to zeta potentials have been used[38]. All the constants involved in the evaluation of DLVO force have been enlisted in Table II and individual force components in Table III.

**TABLE II** Constants used in the evaluation of DLVO force.

| | |
|---|---|
| Particle radius($r$) | $53 \times 10^{-9}$ m |
| Number of particles in closed packed arrangement($n$) | $6$[43] |
| Minimum separation distance between the particles and substrate($z$) | $0.4 \times 10^{-9}$ m [41,42] |
| Permittivity of the fluid medium($\varepsilon$) | $7 \times 10^{-10}$ Fm |
| Reciprocal of the Debye length($k$) | $430 \times 10^{-9}$ m$^{-1}$ |

**TABLE III** Individual force component variation with surface wettability

| Parameter | Plasma treated (hydrophilic) | Glass | PDMS |
|---|---|---|---|
| $A_{123}$ | $-1.64 \times 10^{-20}$ [46] | $3 \times 10^{-20}$ [42] | $5.89 \times 10^{-21}$ [47] |
| $\phi_2$ | $-30$ [49] | $-40$ [45] | $-89$ [44] |
| $F_{VdW, particle-substrate}$ | $-(5.4 \times 10^{-9})$ | $9.863 \times 10^{-9}$ | $1.66 \times 10^{-9}$ |
| $F_{VdW, particle-particle}$ | $1.6562 \times 10^{-9}$ | $1.6562 \times 10^{-9}$ | $1.6562 \times 10^{-9}$ |
| $F_{elctrostatic\ force, particle-substrate}$ | $-(1.74 \times 10^{-17})$ | $-(3.10 \times 10^{-17})$ | $-(15 \times 10^{-17})$ |
| $F_{electrostatic\ force, particle-particle}$ | $-(6.933 \times 10^{-26})$ | $-(6.933 \times 10^{-26})$ | $-(6.933 \times 10^{-26})$ |
| $F_{total}$ | $-3.74 \times 10^{-9}$ | $11.52 \times 10^{-9}$ | $3.32 \times 10^{-9}$ |

Thereby, the interaction parameters accounting for the change in substrate wettability are individual Hamaker constants as well as the surface potential. It is evident from the order of magnitude of the forces enlisted in Table III that the electrostatic force between the particle-substrate may be neglected with respect to the short ranged attractive Van-der-Waals force. This substantiates the premise that the importance of particle–substrate interaction may be attributed to the short range attractive force. This also implies that the decrease in total attractive force with increase in surface hydrophobicity is a manifestation of the dominance of these short range attractive force[44]. We investigate the feasibility of this premise with current experimental investigations.

## 4. Results

The entire phenomena has three major aspects influenced by substrate wettability

### A. Formation of an inhomogeneous colloidal film

There are two competing mechanisms involved in the formation of a stable inhomogeneous film – Rate of contact line retraction and evaporation. The evaporation phenomenon gets complicated due to the presence of colloidal particles. These particles get confined to the contact line, often termed as 'self-pinning'[45]. It has also been reported that the capillary and Marangoni flows get adjunct by additional circulatory flows triggered by the pinning-depinning events of the nanoparticles[46]. This also leads to the absence of the initial spreading stage in the evaporating colloidal droplet. The time required by the droplet to undergo complete phase change and transform into a completely dried film has been designated as the evaporation time. Colloidal droplets of 1µl volume were allowed to evaporate in each category of substrate. It was found that the evaporation rate varied from *5.5 $m^3/s$ to* (for hydrophobic) *0.833 $m^3/s$* (for hydrophilic).

### B. Compaction front and crack front formation

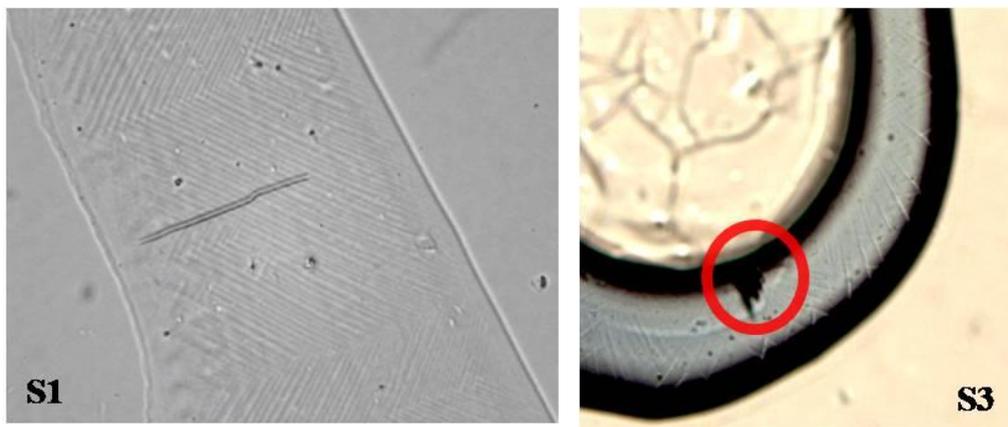

**Fig.1. Compaction front on (a) S1 and (c) S3 – (Advent of Crack front is also shown).**

A visibly dark front is formed due to the evaporation of the solvent and subsequent accumulation of the particles at the edge of the drop (coffee- ring effect)[47]. Area denoted by the dark front in the microsopic image (Fig.1) is better known as the compaction front[48]. Local compaction fronts are formed on substrates

S1 and S2 of similar nature, whereas a uniform solid ring can be observed in case of S3 substrates (Fig.1). Path followed by the crack as it penetrates through the compacted region is called as the crack front[49]. It can be deduced from this two-fold observation that decrease in evaporation rate in turn implies higher residence time of the particles at the edge and it is physically manifestated in the form of distinct compaction fronts.

### C. Crack Dynamics

The entire sequences of a representative crack for each characteristic wettability substrate are shown in Fig.2. Cracks initiate simultaneously at several locations on the droplet edge. Thereby it is practically impossible to rely only on the interrogation window of the recorded video and ascertain the exact time instant of crack initiation. To explain the crack dynamics several numbers of cracks are chosen from the recorded videos which encompass all the three stages mentioned above. Each crack-tip is then followed from its advent till its observable growth ceases. The distance traversed by the crack is termed as the crack propagation. Spatial plots of each crack are differentiated at suitable time intervals to obtain the instantaneous velocity of crack propagation. It has been observed that the crack formation is a highly random process and thereby curves shown in Fig.3 represent a particular trend only.

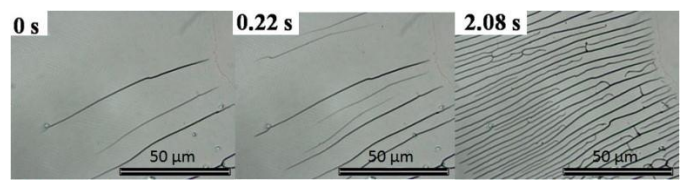

(a)

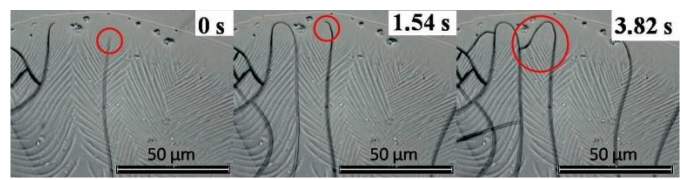

(b)

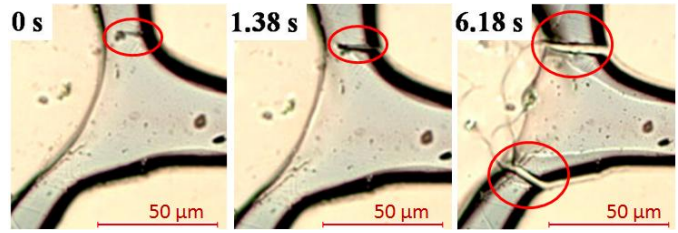

(c)

**Fig.2 Representative Crack formation sequence on (a) S1 (b) S2 (c) S3 substrates**
(Movies in Supporting Information).

Evaporation induced subsequent solidification of particles generates capillary stresses in thin films. This stress maybe approximated as $\sigma = \dfrac{2\gamma}{r}$, numerically equal to 33 MPa for 0.053μm particle diameter. The film tries to expand laterally to relieve this inherent capillary stress. However this expansion is opposed by the lowermost layer of particles bound to the substrate, preventing relaxation of the generated stress. Thereby the only alternative available with the film is to dissipate the stress through creation of cracks[50]. Advancement of the crack through a saturated fluid-solid network necessitates the expenditure of elastic energy. Assuming the film to be thin and elastic, the energy ($E'$) consumed for single crack to progress by an infinitesimal length $dl$ is $\alpha\sigma^2 h^2 \dfrac{dl}{E}$ where $h$ is the film thickness, $E$ is the elastic modulus of the film and $\alpha$ represents the adhesion between the film and substrate ($\alpha$ is assumed to be 1.25 for films attached on only side[51] This energy is consumed to drive the solvent flow as well as to create new surfaces. The instantaneous velocity ($u$) can be obtained as $\dfrac{\alpha\sigma^2 h^2}{E}\dfrac{dl}{dt} \sim \dfrac{\alpha\sigma^2 h^2}{E} u$. In the initial stages the crack tip advances rapidly till it reaches the compaction region. Consequently the maximum stress dissipation is accompanied with viscous losses at the tip. However, the flow gets limited through the compact particle network, arresting crack growth.

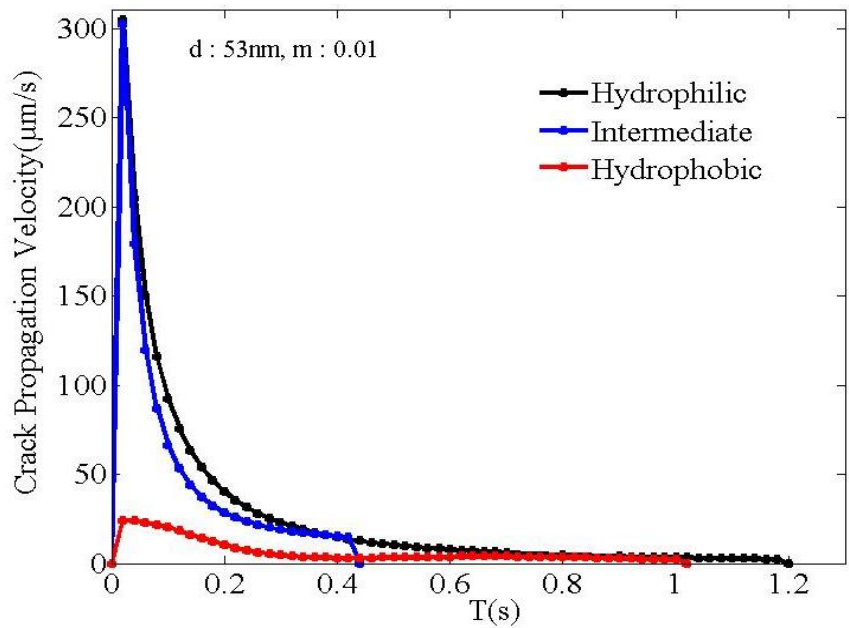

**Fig.3.Effect of surface wettability on crack propagation dynamics**

The characteristic instantaneous velocity is found to lie within the symbolized range (Fig.4). Thereby, the experimentally obtained regimes of crack dynamics corroborate with the previously proposed hypothesis comprising of three distinct regimes- crack initiation, propagation and arrest. The peak propagation velocity for hydrophilic and intermediate substrates are found to be of the order of 300μm/s. However a significant reduction in the peak propagation velocity is observed for hydrophobic substrates (25μm/s)(Fig.3).

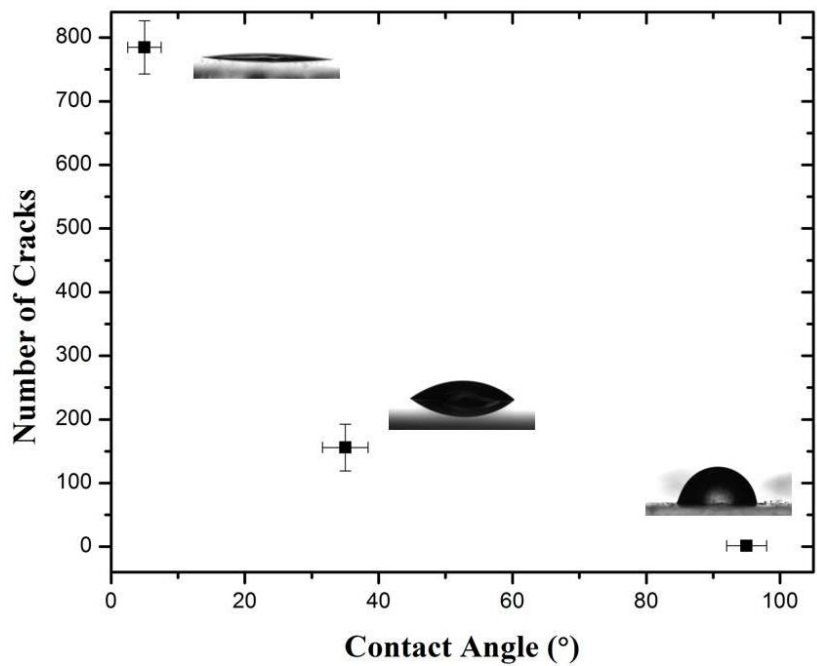

**Fig.4 Number of cracks plotted as a function of contact angle.**

The elastic energy expenditure for crack formation is converted into the bulk surface energy of the crack surface $\Gamma hdL$ where $\Gamma$ [52] is the surface energy/unit crack surface area. However, the prevailing investigation involves conquering an additional resistance offered by the surface energy of the substrate. Hydrophobic surfaces have low surface energy making it easier to rupture the film with an order lower propagation velocity. Though, it follows from the above relation that crack propagation through thicker films would lead to greater loss of energy. Implying an increase in hydrophobicity ought to decelerate crack propagation inspite of being thicker than ther hydrophilic counterparts. This may be attributed to the piece of evidence provided by the reduction in number of cracks(Fig.4) with an increase in substrate hydrophobicity. It indicates that the loss in crack propagation pace is adequately balanced by fewer cracks since the volume (surface area × thickness ) of new surface is generously proportioned.Moreover,the cracks appear to have opened up. This opening up can be perceived as another medium of stress dissipation albeit in a normal direction and is a distinctive feature observed only in case of hydrophobic substrates.These widely opened cracks also indicate lesser lateral shift of consolidated particles as compared to hydrophilic substrates. It may be attributed to the strong particle-substrate repulsion. Channeling this repulsion may assist crack arrest leading to lesser number of cracks and in turn leading to durable film layers.

### D. Morphology and Particle Arrangement

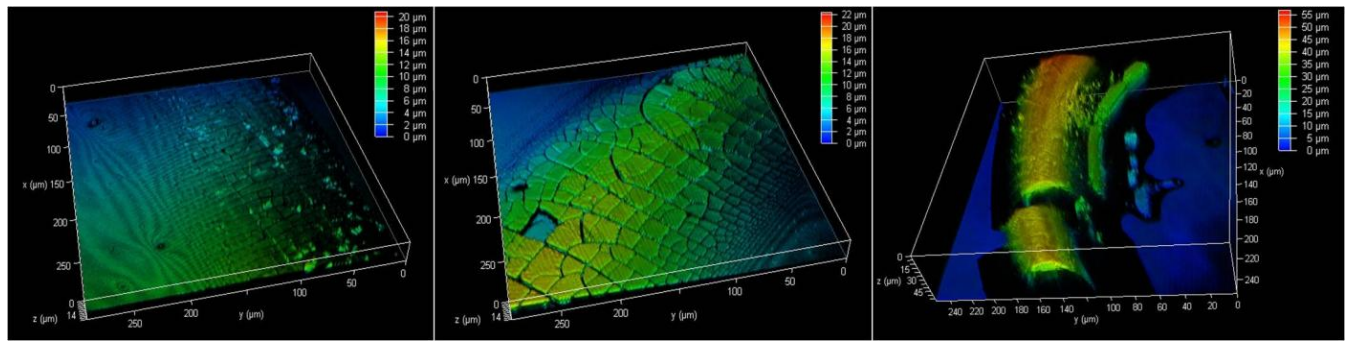

Fig.5  Crack Morphology (L to R) (a) S1 (b) S2 (c) S3

Final stage of crack formation is designated by the point where the emergence of new cracks ceases to occur. The collections of cracks formed thereafter have been classified on the basis of the repetitive structure that represents all the characteristics of the pattern as a whole. Fig.5 shows the representative confocal images of the cracks where in numerous fine hair-like cracks, can be seen to propagate on hydrophilic substrates from the edge of the drop to the centre simultaneously. Secondary cracks on the other hand can be observed to propagate from previously formed primary cracks. The growth of the secondary cracks is limited by the distance of the primary cracks. Primary cracks on the other hand get readily coupled with the secondary cracks to give rise to a network structure. Intermediate substrates on the other hand gave rise to generic arch shaped[27] cracks that are usually found in dried nanolatex films. Each arch can be designated as the parent crack whereas daughter cracks originate across the length and breadth. These daughter cracks divide the parent crack into small compartments. The difference in the instantaneous distance between the gelation/compaction front and the crack tip is said to be responsible for the observed characteristic morphology. It also corroborates the claim of 'difficulty in nucleation'[27] in nanolatex gels held responsible for the formation of this characteristic pattern. 'Trench' cracks are observed on hydrophobic substrates where in each crack forms a deep trough and its propagation is stalled within a certain distance from the edge of the drop unlike radial cracks which travel right up to the droplet center.

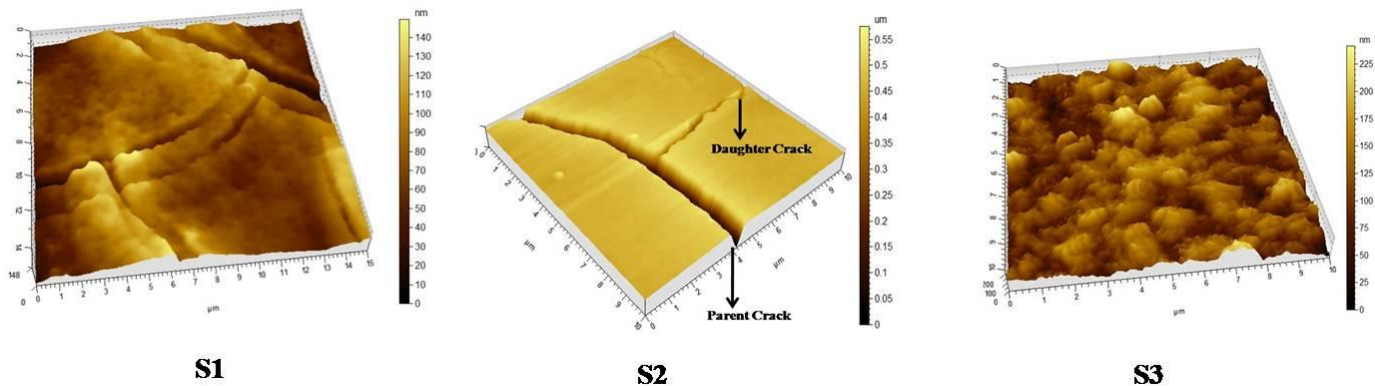

**Fig.6. AFM images of crack morphology**

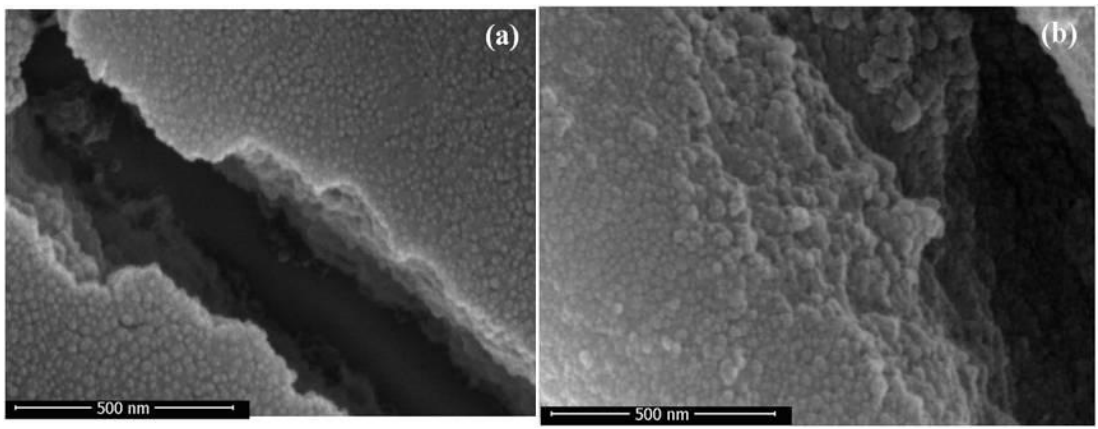

**Fig.7 SEM images of crack morphology (a) S1 (b) S2**

To unravel the microstructure beneath such morphologically different cracks, we resorted to AFM (Fig.6) and SEM imaging (Fig.7). The details of the imaging technique have been provided in the experimental section. We focus at the areas in the vicinity of the crack. The area near the crack of S2 is scanned (3 μm x 3 μm.) and its corresponding phase image shows compact particle arrangement. The AFM tip could not reach the bottom of crack in case of a hydrophobic substrate thereby particles could not be distinguished individually[8], though islands of particle agglomerates have been observed[53]. This suggests that particles prefer coupling with each other than to the substrate indicating that particle-substrate repulsion induces such arrangement. Thereby strength of compaction also reflects the stress experienced by the particles.

Consequently, it may be proposed that substrate wettability manipulation may be one of the keys to control final film morphology and variants (hair-like, arch and trench) of cracks may be obtained by altering a single variable-surface wettability.

**E. Influence of substrate wettability on stress dissipation:** Crack advancement is closely linked with the local stress developed in the colloidal films and its propagation is directed by the local stress field and film properties. This stress variation is in turn characterized by a stress intensity factor *K* and a critical stress intensity factor *Kc*, measure of dried film's resistance to rupture. The morphologically different cracks are similar in their mode of opening (mode-I)[54]. Thereby it is possible to characterize the cracks on the basis of *crack opening δ* and *the distance along the crack from its tip r*, related for mode-I (opening mode) crack as [54]

$$\delta = \frac{8Kc}{E} \sqrt{\frac{r}{2\pi}}$$

The equation relates the experimentally measureable parameters with the critical stress intensity factor. It has been assumed that the film is elastic. Moreover, the regions very close to the crack tip bear permanent damage caused by the stress. Thereby the applicability of the equation stands valid at regions where it is capable of overcoming the deformation thoroughout the given film thickness[55,51]. Physically, the Young's modulus of the film remains unchanged since it is a material property, independant of surface characteristics. The stress at the crack tip is null and increases progressively as we move away from the crack tip. We have measured these parameters, δ and *r* for several sets of cracks on each type of substrate as shown in Fig.8

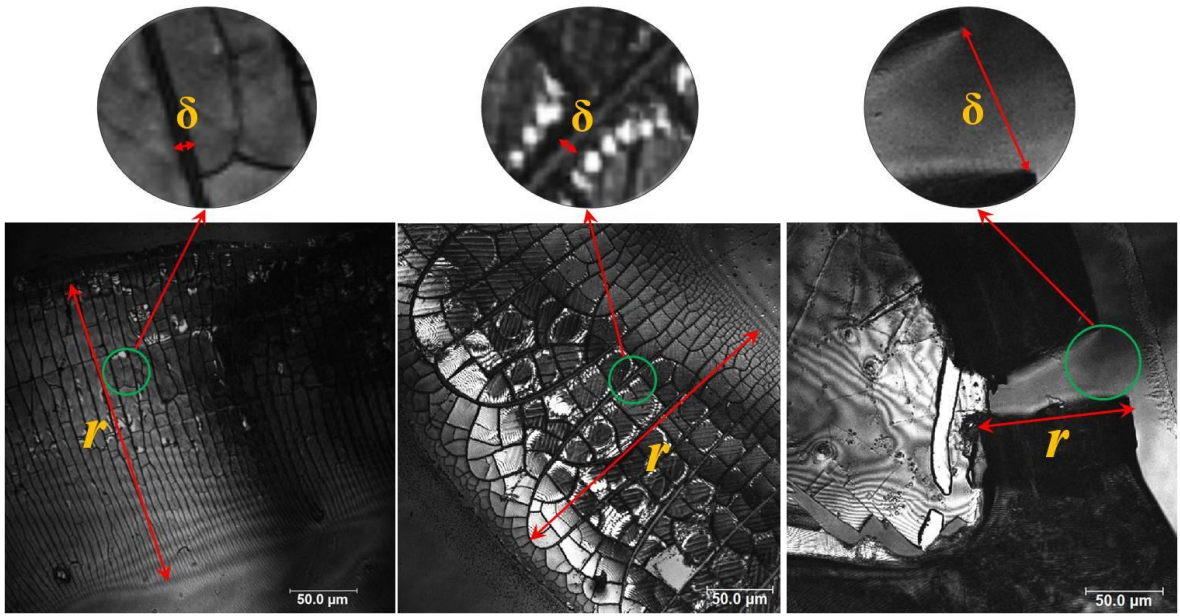

**Fig.8. Evaluation of Crack Opening δ and distance along the crack from its tip r**

**(L to R) (a) S1 (b) S2 (c) S3**

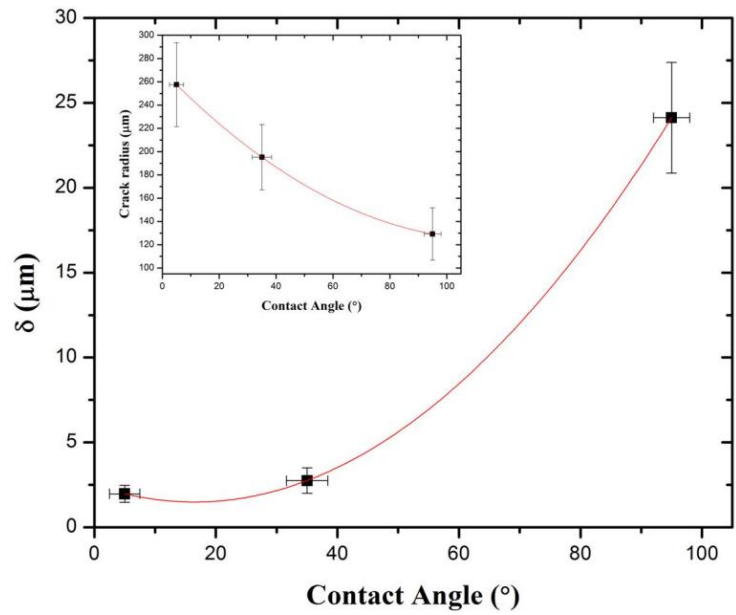

**Fig.9. Effect of wettability on crack characteristics**

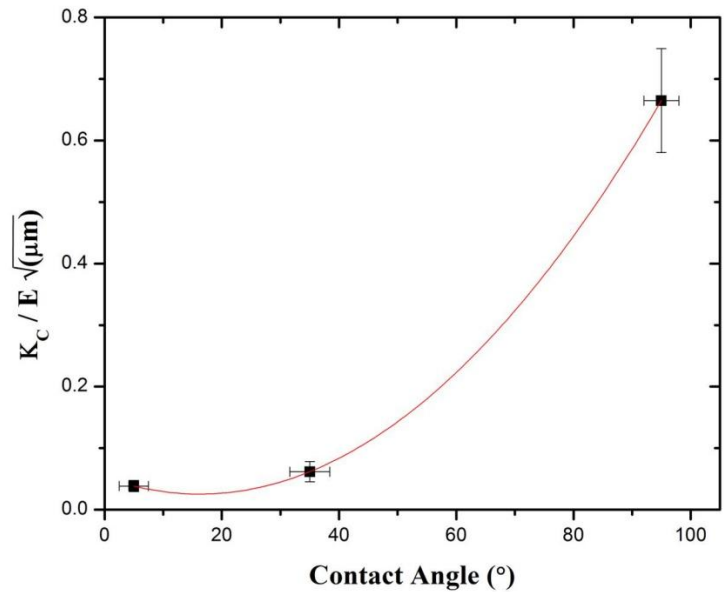

**Fig.10. Effect of wettability on stress intensity factor**

Fig.9 clearly shows that the crack opening increases with surface hydrophobicity inspite of the decreasing distance from the crack tip.This can be interpreted as the response of the system (film) to the magnitude of developed internal stresses. It is known that a film will undergo rupture only if $K \geq K_c$ and it follows from Fig.10 that the critical stress intensity factor increases with substrate hydrophobicity. This suggests that the films on hydrophobic substrates have a higher threshold stress limit leading to better durability. We have already deduced that the capillary stresses generated internally is also of a higher order for hydrophobic substrates. Moreover, it has been reported that the stress in a material under confined configurations shows an increase proportional to material thickness. This is where hydrophobic films gain an advantage, they are not only thicker but they are also capable of enduring higher stress. Experimental evidence marking the point of rupture is manifested in the dynamic delay in crack initiation subsequent to compaction front formation. We have reported experimentally that distinct compaction fronts are formed on hydrophobic substrates as compared to local compaction on hydrophilic substrates. This is attributed to the preference of particle-particle interaction as compared to particle-substrate interactions in the former.Fig.1 also corroborates this experimental finding since this local compaction fronts are initiated at several locations due to the incompetence of hydrophilic films in bearing the stress.

**Conclusions**

The functionality of substrate surface wettability on crack formation process in colloidal films has been experimentally investigated. Microdroplets of colloidal naosuspensions have been subjected to natural drying on substrates with varying surface energies. The aspect specifically influenced by substrate wettability is the substrate-particle interaction governed by the total attractive force as stated by the DLVO theory. Substrate-particle interactions are represented by the magnitude of Hamaker constant and is duly accounted for in the Van-der-Waals force whereas the effect of electrostatic repulsion has been captured as a function of the surface potential.

Hydrophobic substrates (25μm/s) have shown a marked decrease in crack propagation velocity as compared to their hydrophilic (300μm/s) counterparts. We also characterize the wettability induced morphologically different cracks using optical and confocal microscopy techniques. The number of cracks is found on undergo a drastic reduction with increase in surface hydrophobicity. This has been adequately supported by the critical intensity factor depicting its capacity to bear stress. The other facets of surface characteristics of substrate should be investigated in detail. Future research will explore the effect of particle interactions on crack dynamics.